\documentclass[conference]{IEEEtran}

\IEEEoverridecommandlockouts
\def\BibTeX{{\rm B\kern-.05em{\sc i\kern-.025em b}\kern-.08em
		T\kern-.1667em\lower.7ex\hbox{E}\kern-.125emX}}

\usepackage{dblfloatfix}    

\usepackage{cite}
\usepackage{amsmath,amssymb,amsfonts}
\usepackage{bbm}

\usepackage{graphicx}
\usepackage{textcomp}
\usepackage{xcolor}
\usepackage{colortbl} 

\usepackage{url} 

\usepackage{bm}
\usepackage[normalem]{ulem} 
\usepackage{supertabular}
\usepackage{amsthm}
\theoremstyle{definition}

\newtheorem{prop}{Proposition}
\usepackage[linesnumbered,ruled,vlined]{algorithm2e}
\usepackage{algorithmicx} 
\usepackage{pbox}
\usepackage{float} 
\usepackage{subcaption} 
\usepackage{boldline,multirow}
\usepackage{array}

\usepackage{pbox}

\usepackage{dblfloatfix}

\newcounter{probNum}

\begin{document}

%
\title{Resource Allocation and Scheduling in Non-coherent User-centric Cell-free MIMO}
%
%
\author{Hussein~A.~Ammar\IEEEauthorrefmark{1}, Raviraj~Adve\IEEEauthorrefmark{1}, Shahram~Shahbazpanahi\IEEEauthorrefmark{2}\IEEEauthorrefmark{1}, Gary~Boudreau\IEEEauthorrefmark{3}, and~Kothapalli~Srinivas\IEEEauthorrefmark{3}
	\\
	\IEEEauthorrefmark{1}University of Toronto, Dep. of Elec. and Comp. Eng., Toronto, Canada\\
	\IEEEauthorrefmark{2}University of Ontario Institute of Technology, Dep. of Elec. and Comp. Eng., Oshawa, Canada\\
	\IEEEauthorrefmark{3}Ericsson Canada, Ottawa, Canada\\
}


%
%

\maketitle 


\newcommand*{\myResultSELow}{$8$} 
\newcommand*{\myResultSEHigh}{$10$}
\newcommand*{\myResultOne}{$9.1$}
\newcommand*{\myResultTwo}{$10.6$}
\newcommand*{\myResultThree}{$1.67$}
\newcommand*{\myResultFour}{$5.44$}
\newcommand*{\myResultReuseFactor}{$0.32$} 
\newcommand*{\myResultReuseFactorTwo}{$0.16$} 
\newcommand*{\myResultReuseFactorThree}{$0.08$} 
\newcommand*{\myResultReuseFactorPerf}{$65.64$} 
\newcommand*{\myResultReuseFactorPerfTwo}{$70.8$} 
\newcommand*{\myResultReuseFactorPerfThree}{$78.83$} 
\newcommand*{\myResultReuseFactorPerfNF}{$49.47$} 
\newcommand*{\myResultReuseFactorPerfNFTwo}{$65.24$} 
\newcommand*{\myResultReuseFactorPerfNFThree}{$77$} 
\newcommand*{\myResultReuseFactorPerfRobust}{$39.21$} 
\newcommand*{\myResultReuseFactorPerfRobustTwo}{$37$} 
\newcommand*{\myResultReuseFactorPerfRobustThree}{$43.21$}
\newcommand*{\myResultReuseFactorPerfRobustNF}{$10.6$} 
\newcommand*{\myResultReuseFactorPerfRobustNFTwo}{$25$} 
\newcommand*{\myResultReuseFactorPerfRobustNFThree}{$38.27$}
\newcommand*{\myResultReuseFactorPerfRobustSingleTS}{$37$} 
\newcommand*{\myResultReuseFactorPerfRobustSingleTSTwo}{$28.62$} 
\newcommand*{\myResultReuseFactorPerfRobustSingleTSThree}{$32.1$} 
\newcommand*{\myResultReuseFactorPerfRobustSingleTSNF}{$7.32$} 
\newcommand*{\myResultReuseFactorPerfRobustSingleTSNFTwo}{$15.02$} 
\newcommand*{\myResultReuseFactorPerfRobustSingleTSNFThree}{$26.18$} 

\begin{abstract}
	We study the problem of user-scheduling and resource allocation in distributed multi-user, multiple-input multiple-output (MIMO) networks implementing user-centric clustering and non-coherent transmission. We formulate a weighted sum-rate maximization problem which can provide user proportional fairness. As in this setup, users can be served by many transmitters, user scheduling is particularly difficult. To solve this issue, we use block coordinate descent, fractional programming, and compressive sensing to construct an algorithm that performs user-scheduling and beamforming. Our results show that the proposed framework provides an \myResultSELow-\ to \myResultSEHigh-fold gain in the long-term user spectral efficiency compared to benchmark schemes such as round-robin scheduling. Furthermore, we quantify the performance loss due to imperfect channel state information and pilot training overhead using a defined \emph{area-based} pilot-reuse factor.
\end{abstract}
\begin{IEEEkeywords}
	User-centric clustering, cell-free, user-scheduling, resource allocation, distributed MIMO, distributed antennas system, fairness, imperfect CSI.
\end{IEEEkeywords}

%

%
%
%
\section{Introduction}
Deploying user-centric clustering in distributed multiple-input multiple-output (MIMO) networks enhances the performance of the conventional cell-edge users by placing each user at the effective center of its serving cluster~\cite{PDPUsercentricVsDisjoint8969384, cellFreeVersusSmallCells7827017}. User-centric clustering can outperform general cell-free networks that assume all the remote radio heads (RRHs) can serve the users~\cite{cellFreeUserCentricPower8901451}. 
Recently, resource allocation under cell-free MIMO has attracted significant attention. The studies in~\cite{PrecodingDistrib2020Atzeni, 6920005} optimize beamforming design by minimizing a weighted sum mean square error (MSE) utility, which is easier to tackle than weighted sum rate (WSR) maximization problems but suffers a penalty in terms of sum-rate~\cite{PrecodingDistrib2020Atzeni}. 

The work in~\cite{cellFreeUserCentricPower8901451} considers optimizing power allocation to maximize lower bounds for sum-rate and minimum rate. Similarly,~\cite{maxMinRate8756286} optimizes the beamforming to maximize the minimum rate. Note that, max-min rate solutions do not provide flexibility to control the fairness. Moreover, the authors in~\cite{powerControlCellFree7917284} consider a near-optimal power control algorithm using zero-forcing (ZF) and conjugate beamforming that is simpler than the max–min power approach for cell-free massive MIMO networks. Furthermore, the work in~\cite{9107496} optimizes the beamforming by using a lower-bound for the logarithm function of the rate to obtain a local optimum.

A crucial component in optimizing the WSR is user-scheduling. In conventional networks, techniques that have been investigated include dual decomposition and the gradient method~\cite{ResourceAllo6175089}, where the scheduling variables are relaxed from being binary. Notably, this relaxation is optimal for a large number of subcarriers~\cite{ResourceAllo1658226}. 
Furthermore, the investigations in~\cite{FR8310563, Ahmad9084256} use fractional programming to perform resource allocation in conventional networks, where the user-scheduling part is performed using a combinatorial search. 



In summary, the main limitation of these studies is either not specifically addressing the user-centric clustering scheme, or ignoring the user-scheduling step for the users, that is, the scheduled users are assumed to be \emph{preselected}. In this paper, we optimize user-scheduling and resource allocation in a user-centric cell-free MIMO network through formulating a WSR problem. We study the non-coherent transmission mode, which does not require the RRHs to strictly synchronize their transmissions, but it prevents from directly using the weighted minimum mean square error (WMMSE)~\cite{NoncoherentCRAN8482453}. 
The scheduling part of the problem cannot be solved efficiently using a combinatorial search algorithm because each user can be served by many RRHs with overlapping serving clusters. To tackle this, we employ tools from block coordinate descent, fractional programming, and compressive sensing, which allow the construction of an algorithm that guarantees convergence of the network sum-rate through a smooth non-decreasing pattern.  In summary, the contributions of this paper are:
\begin{itemize}
	\item Formulating the WSR problem for the non-coherent cell-free multiuser MIMO setting
	\item Using fractional programming to optimize beamforming and employing compressive sensing to solve the scheduling problem
	\item Developing and implementing robust beamforming to account for channel estimation errors
\end{itemize}

The rest of the paper is organized as follows. Section~\ref{section:Model} presents our system model, while Section~\ref{section:prob_formulation} formulates the optimization problem. Section~\ref{section:ResourceAllocation} presents our proposed resource allocation algorithm. Finally, Sections~\ref{section:results} and~\ref{section:conclusion} report our simulation results and conclusion, respectively.

\section{System Model}\label{section:Model}
\subsection{Network Model}
\begin{figure}
	\vspace{-0.5em}
	\centering
	\includegraphics[width=0.8\linewidth]{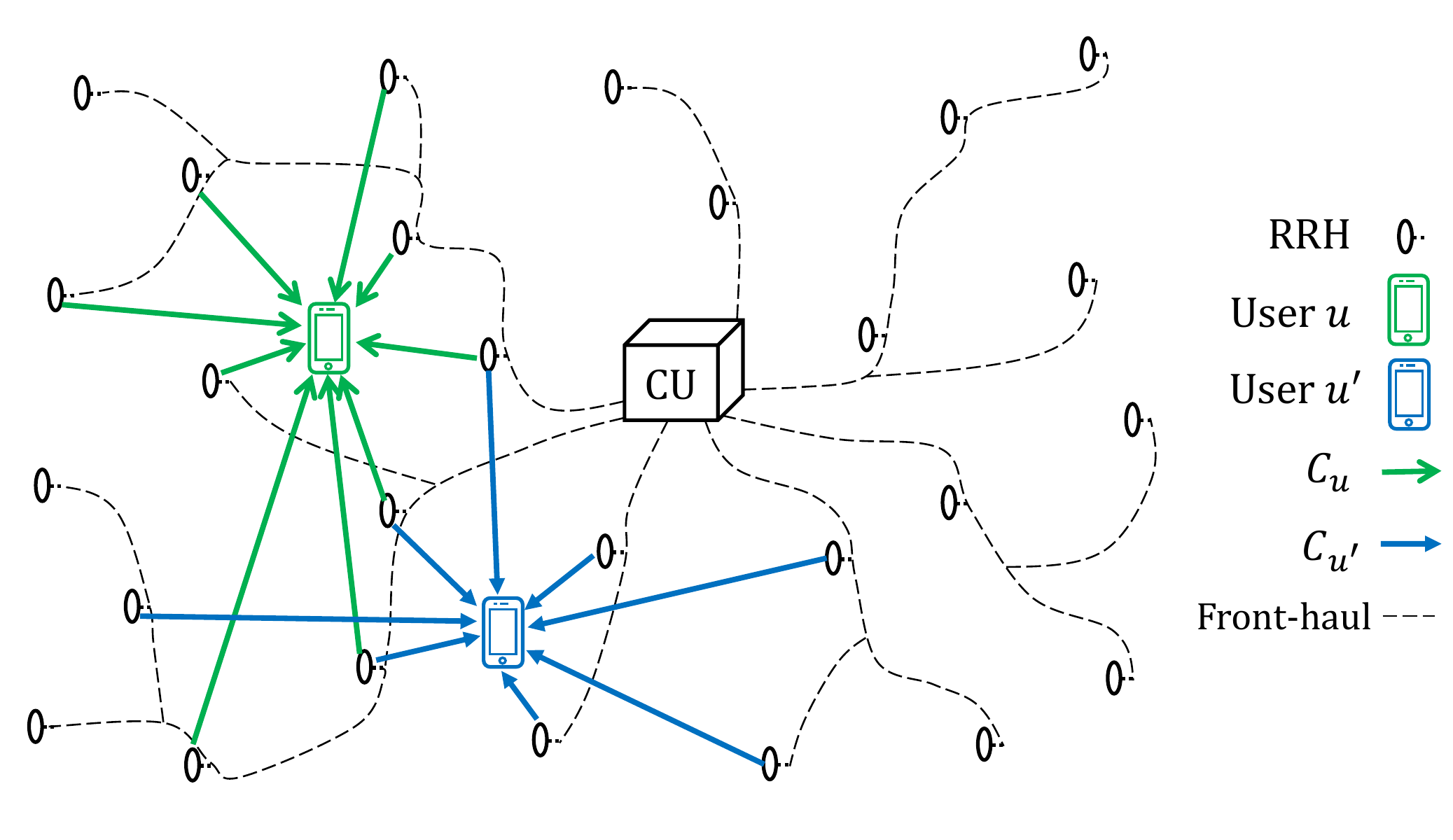}
	\caption{Serving cluster using user-centric clustering.}
	\label{fig:UserCentricClustering}
	\vspace{-1em}
\end{figure}

As shown in Fig.~\ref{fig:UserCentricClustering}, we consider the downlink of a time-division duplex (TDD) system comprising several RRHs, represented by the set $\mathcal{B}$, each equipped with $M$ antennas and jointly serving the active users, represented by the set $\mathcal{U}$. Both RRHs and users are randomly located in 2D space. The RRHs are controlled by a single control unit (CU), and as in~\cite{cellFreeUserCentricPower8901451}, we assume a relaxed front-haul constraint, which can be realized through technologies like the radio stripes system~\cite{frenger2019antenna}.

For each user $u \in \mathcal{U}$, we define a cluster $\mathcal{C}_u$ that includes the RRHs that \emph{potentially can be selected} to serve the user according to user-centric clustering. Specifically, $\mathcal{C}_u$ comprises the RRHs with strong average channels, i.e., $\mathcal{C}_u = \{ r \mid \left(\psi_{ru}L(d_{ru}) \right) \ge \rho\}$, where $\psi_{ru}$ denotes the shadowing, $L(d_{ru})$ accounts for the path loss; here, $d_{ru}$ is the distance between RRH $r$ and user $u$. If no RRH meets this criterion, the $\mathcal{C}_u$ for the user comprises the RRH with largest $\left(\psi_{ru}L(d_{ru}) \right)$. Finally, we represent the users that may be served by RRH $r$ as $\mathcal{E}_r$. 

\subsection{Channel Estimation}
Channel estimation is performed through an uplink pilot-training phase of length $\tau_p$. During this phase, we can write the signal ${\bf Y}_r \in \mathbb{C}^{M \times \tau_p}$ received at RRH $r$~as
\begin{align}
{\bf Y}_r = \sum_{u \in \mathcal{U}} \sqrt{p_u} {\bf h}_{ru} {\bm \Phi}_u + {\bf Z}_r,
\end{align}
where ${\bm \Phi}_u \in \mathbb{C}^{1 \times \tau_p}$ is the unit norm (${\bm \Phi}_u {\bm \Phi}_u^H = 1$) pilot sequence used by user $u$, $p_u$ is the transmit power of the user, and ${\bf Z}_r$ is the additive white Gaussian noise (AWGN) with entries $\sim \mathcal{CN}\left(0, \sigma_Z^2 \right)$; ${\bf h}_{ru} \in \mathbb{C}^{M \times 1}$, the channel between RRH $r$ and user $u$ is modeled as ${\bf h}_{ru} \triangleq \sqrt{\psi_{ru} L(d_{ru})} {\bf g}_{ru}$, where ${\bf g}_{ru} \sim \mathcal{CN}({\bf 0},{\bf I}_M)$ accounts for small-scale fading.

As in~\cite{cellFreeVersusSmallCells7827017}, we assume knowledge of the users' transmit powers and large-scale fading. Hence, using ${\bf \breve{y}}_r = \mathrm{vec}\{{\bf Y}_r\} \in \mathbb{C}^{M \tau_p  \times 1}$ and linear MMSE, the channel estimate ${\bf \hat{h}}_{ru}, \forall u \in \mathcal{E}_r$ can be obtained as ${\bf \hat{h}}_{ru} = {\bf R}_{ru} {\bf R}_{r}^{-1} {\bf \breve{y}}_r,$
with ${\bf R}_{ru} = \sqrt{p_u} \psi_{ru} L(d_{ru}) \left( {\bm \Phi}_u^* \otimes {\bf I}_M \right)$ and 
\[
{\bf R}_{r} = \sum_{u \in \mathcal{U}} p_u \psi_{ru} L(d_{ru}) \left( {\bm \Phi}_u^T {\bm \Phi}_u^* \otimes {\bf I}_M \right) + \sigma_z^2 {\bf I}_{M \tau_p}.
\]

When the number of users $|\mathcal{U}| \ge \tau_p$, the available pilot sequences need to be reused by the users, adding pilot contamination. This results in the estimated channel ${\bf \hat{h}}_{ru} \sim \mathcal{CN}\left( {\bf 0}, {\bm \Psi}_{ru} \right)$, with the error covariance matrix given by~\cite{kay1993fundamentals}
\begin{align}
{\bm \Psi}_{ru} \triangleq {\bf D}_{ru} \left( \sum_{u' \in \mathcal{U}_u} {\bf D}_{ru'} + \frac{\sigma_Z^2}{p_u} {\bf I}_M \right)^{-1} {\bf D}_{ru},
\end{align}
where ${\bf D}_{ru} \in \mathbb{C}^{M \times M}$ is a diagonal matrix with diagonal entries $\left[{\bf D}_{ru}\right]_{mm} \triangleq \psi_{ru} L(d_{ru})$, and $\mathcal{U}_u$ is the set of users employing the same pilot sequence as user $u$ (including user $u$). It is known from MMSE that the channel estimation error $\mathrm{ {\bf e}}_{ru} = {\bf h}_{ru} - {\bf \hat{h}}_{ru}$ is uncorrelated with ${\bf \hat{h}}_{ru}$ and can be modeled as $\mathrm{ {\bf e}}_{ru} \sim \mathcal{CN}\left({\bf 0}, {\bm \Theta_{ru}}\right)$, where ${\bm \Theta_{ru}} \triangleq {\bf D}_{ru} - {\bm \Psi}_{ru}$.

The downlink signal received at user $u$ can be modeled as
\begin{align}\label{eq:signalModel_imperfectChan}
&y_{u}
=
\sum_{r \in \mathcal{C}_u} \sqrt{{s}_{ru}}
\left( {\bf \hat{h}}_{ru}^H + \mathrm{ {\bf e}}_{ru}^H \right) {\bf w}_{ru} x_{ru}
\nonumber \\
& \ 
+
\resizebox{0.41\textwidth}{!}
{$\displaystyle
	\sum_{r' \in \mathcal{B}} \sum_{u' \in \mathcal{E}_{r'},u' \ne u} \sqrt{{s}_{r'u'}} \left( {\bf \hat{h}}_{r'u}^H + \mathrm{ {\bf e}}_{r'u}^H \right) {\bf w}_{r'u'} x_{r'u'}
	+ z_{u}
	$}
\end{align}
where $\{x_{ru}: r \in \mathcal{C}_u\}$ are the symbols transmitted by the serving RRHs for user $u$ with $\mathbb{E}\{|x_{ru}|^2\} = 1$, ${\bf w}_{ru} \in \mathbb{C}^{M \times 1}$ is the precoding vector used by RRH $r$ to serve user $u$, 
and $z_{u} \sim\mathcal{CN}(0, \sigma_z^2)$ is the AWGN.

\subsection{Pilot Assignment (PA) Policy}
Properly assigning the pilots to the users is clearly pivotal to decrease pilot contamination.

\begin{prop}\label{prop:clustering}
	We propose to use a heuristic low-overhead location-based PA policy. Our policy assigns non-orthogonal pilots for users that are far from each other by using the hierarchical agglomerative clustering (HAC) algorithm~\cite{karypis2000comparison}. 
	The HAC creates a tree to cluster the users into many groups each containing a number of users less than or equal to the number of available orthogonal pilot sequences. We then assign each group the available orthogonal sequences. The algorithm can be constructed as follows:
	\begin{enumerate}
		\item Treat each active user as a cluster head.
		\item Combine the two nearest clusters into one using an average linkage, e.g., Ward's minimum variance criterion.
		\item Repeat Step 2 until you reach the root of the tree where all the users are in the same cluster.
		\item While backtracking the tree starting from the root, define each cluster when its number of users is less than or equal $\tau_p$.
		\item Assign the orthogonal pilots to each cluster randomly.\label{step:HAC_assignP}
	\end{enumerate}
\end{prop}
The HAC algorithm is more consistent than the K-means and Gaussian mixture models, and it is not sensitive to the choice of the used distance-metric~\cite{karypis2000comparison}.
Also, as this algorithm does not require selecting the number of clusters needed, it allows us to easily define the cluster based on an upper limit of the number of users belonging to it, i.e., relate it to $\tau_p$. 

\section{Problem Formulation}\label{section:prob_formulation}
\subsection{Problem Definition}
To decode the data streams from the RRHs, the users employ successive interference cancellation (SIC). Under the assumption of perfect SIC, the effective achievable rate\footnote{This expression is based on using the famous Jensen's Inequality to write down a lower-bound for the data rate with an expectation over the unknown instantaneous channel state information (CSI) error $\{\mathrm{ {\bf e}}_{ru}: r \in \mathcal{B}, u \in \mathcal{E}_r\}$, i.e., $\mathbb{E}_{\mathrm{ {\bf e}}}\left\{\log\left(1 + \widetilde{\gamma}_{u} \right) \right\} \ge \log\left(1 + 1/\mathbb{E}_{\mathrm{ {\bf e}}}\left\{\widetilde{\gamma}_{u}^{-1} \right\} \right)$, then using SIC to decode the received data streams at the user. Note that this expression is only used to perform the resource allocation, however, when we plot the performance, we use the actual achievable rate using the actual channels.} for user $u$ can be modeled by the CU as~\cite{NoncoherentCRAN8482453}
\begin{align}\label{eq:rate}
R_u =
\frac{\left(\tau_d - \tau_p\right)} {\tau_d}
\log\Bigg(1 + \frac{ 
	\sum_{r \in \mathcal{C}_u} {s}_{ru}
	| {\bf \hat{h}}_{ru}^H {\bf w}_{ru} |^2 }
{
	A_{u}\left(\mathcal{S}, \mathcal{W} \right)
} \Bigg),
\end{align}
\vspace{-1em}
\begin{align}\label{eq:simplifyingTerm}
\text{with}\quad
& A_{u}\left(\mathcal{S}, \mathcal{W} \right) =
\sum_{r' \in \mathcal{B}} \sum_{u' \in \mathcal{E}_{r'},u' \ne u} {s}_{r'u'} \left| {\bf \hat{h}}_{r'u}^H {\bf w}_{r'u'} \right|^2
\nonumber \\
& \quad \quad 
+ \sum_{r' \in \mathcal{B}} \sum_{u' \in \mathcal{E}_{r'}}  {s}_{r'u'} {\bf w}_{r'u'}^H {\bm \Theta_{r'u}}  {\bf w}_{r'u'}
+ \sigma_z^2
\end{align}
where $\tau_d$ is the channel coherence time, $\mathcal{S} = \{{\bf s}_1, \dots, {\bf s}_{|\mathcal{B}|}\}$ is the set of binary scheduling variables at the RRHs with ${\bf s}_r = [s_{r u_1}\ \dots\ s_{r u_{|\mathcal{E}_r|}}]^T \in \mathbb{B}^{|\mathcal{E}_r| \times 1}$, i.e, if $s_{r u} = 1$, user $u$ is scheduled by RRH $r$, else it is not. Similarly, the set of the beamformers is $\mathcal{W} = \{{\bf W}_1,\ \dots,\ {\bf W}_{|\mathcal{B}|}\}$ with ${\bf W}_r = [{\bf w}_{ru_1},\ \dots,\ {\bf w}_{ru_{|\mathcal{E}_r|}}] \in \mathbb{C}^{M \times |\mathcal{E}_r|}$. The term ${\bm \Theta_{ru'}} = \mathbb{E}\{ \mathrm{ {\bf e}}_{ru'} \mathrm{ {\bf e}}_{ru'}^H \}$ is the covariance of the estimation error of the channel between RRH $r$ and user $u'$, and including it in the model allows to construct a robust beamforming.

We formulate the following WSR problem on the CU
\begin{subequations}\label{eq:TotalProblem_1}
	\begin{align}
	\stepcounter{probNum}
	(\mathrm{P}\arabic{probNum})\quad
	\max_{ \mathcal{S}, \mathcal{W}}\quad & \sum_{u \in \mathcal{U}} \delta_{u} \log\left( 1 + 
	\gamma_{u}
	\right) 
	\label{eq:TotalProblem_1_obj}
	\\
	\text{s.t.}\quad 
	& \sum_{u \in \mathcal{E}_r} {s}_{ru} \le M,
	\mkern130mu
	r \in \mathcal{B}
	\label{eq:TotalUtilityProblem_U_1}
	\\
	&
	\sum_{u\in \mathcal{E}_r} \|{\bf w}_{ru}\|_2^2 \le p,
	\mkern105mu
	r \in \mathcal{B}
	\label{eq:TotalUtilityProblem_U_1_w}
	\\
	&
	\gamma_{u} = \frac{ 
		\sum_{r \in \mathcal{C}_u} {s}_{ru}
		\left| {\bf \hat{h}}_{ru}^H {\bf w}_{ru} \right|^2 }
	{
		A_{u}\left(\mathcal{S}, \mathcal{W} \right)
	},
	\mkern25mu
	u \in \mathcal{U}
	\label{eq:TotalUtilityProblem_U_1_SINR}
	\\
	&
	{s}_{ru} \in \{0, 1\}
	\mkern100mu
	r \in \mathcal{B}, u \in \mathcal{E}_r
	\label{eq:TotalUtilityProblem_U_1_tau}
	\end{align}
\end{subequations}
where $\delta_{u}$ denotes the proportional fair weights for user $u$. The term $A_{u}$ is defined in~\eqref{eq:simplifyingTerm}. Problem~\eqref{eq:TotalProblem_1} optimizes the decision variables $\mathcal{S}$ and $\mathcal{W}$ which determine the user-scheduling and beamforming weight vectors, respectively, such that the total utility in~\eqref{eq:TotalProblem_1_obj} is maximized. We ignore the pre-log pilot training overhead factor because it is a constant. 
Constraints~\eqref{eq:TotalUtilityProblem_U_1} prevent the RRHs from simultaneously serving more than $M$ users on the same channel. Constraints~\eqref{eq:TotalUtilityProblem_U_1_w} satisfy the power budget of the RRHs, and~\eqref{eq:TotalUtilityProblem_U_1_tau} show that a user $u$ can be scheduled or not. Constraints~\eqref{eq:TotalUtilityProblem_U_1_SINR} define the effective signal to interference and noise ratio (SINR) as an auxiliary variable.

Problem~\eqref{eq:TotalProblem_1} is a mixed-integer non-convex problem and obtaining a global optimum is mathematically prohibitive.

\subsection{Problem Analysis}\label{sectionc:Analysis}
The beamforming vectors are constructed for users that are actually scheduled on the channel and hence
\vspace{-0.5em} 
\begin{align}
{s}_{ru} = \mathbbm{1} \{ \|{\bf w}_{ru}\|_2^2 \}
=
\left\| \left\| {\bf w}_{ru} \right\|_2^2 \right\|_0
\end{align}
where $\left\| \cdot \right\|_0$ is the $\ell_0$-norm. Using the literature of compressive sensing, the $\ell_0$-norm of a vector ${\bf x}$ can be approximated as a weighted convex $\ell_1$-norm $\|{\bf x}\|_0 \simeq \sum_{m} \alpha_m |x_m| = \| {\bm \alpha} {\bf x}\|_1$~\cite{candes2008enhancing},
where $\alpha_m$ are positive weights that penalize the nonzero coefficients $x_m$, and ${\bm \alpha} = \textbf{diag}\{\alpha_1, \alpha_2, \dots\}$ is a diagonal matrix. For our case, ${\bf x} = \left\| {\bf w}_{ru} \right\|_2$ which is scalar. We can construct an iterative process to find these weights at each iteration $i$ as suggested in~\cite{candes2008enhancing}
\vspace{-0.4em}
\begin{align}\label{eq:weightsUpdate}
\alpha_{ru}^{(i+1)} = \frac{1}{ \left\| {\bf w}_{ru}^{(i)} \right\|_2^2 + \epsilon} \ ,
\end{align}
where $\epsilon > 0$ provides stability and ensures that a zero-valued component in $\left\| {\bf w}_{ru} \right\|_2^2$ does not strictly prohibit a nonzero estimate at the update in the next iteration. 

As a result, our problem can be formulated as follows
\begin{subequations}\label{eq:TotalUtilityProblem_w}
	\begin{align}
	\stepcounter{probNum}
	(\mathrm{P}\arabic{probNum})\quad
	\max_{ \mathcal{W}}\quad & \sum_{u \in \mathcal{U}} \delta_{u} \log\left( 1 + 
	\gamma_{u}
	\right)
	&
	\label{eq:TotalUtilityProblem_w_obj}
	\\
	\text{s.t.}\quad 
	& \sum_{u \in \mathcal{E}_r} \alpha_{ru} \|{\bf w}_{ru}\|_2^2 \le M,
	\mkern70mu
	r \in \mathcal{B}
	\label{eq:TotalUtilityProblem_w_indc1}
	\\
	&
	\sum_{u\in \mathcal{E}_r} \|{\bf w}_{ru}\|_2^2 \le p,
	\mkern105mu
	r \in \mathcal{B}
	\label{eq:TotalUtilityProblem_w_powerBudget}
	\\
	&
	\gamma_{u} = \frac{ 
		\sum_{r \in \mathcal{C}_u}
		\left| {\bf \hat{h}}_{ru}^H {\bf w}_{ru} \right|^2 }
	{
		B_{u}\left( \mathcal{W} \right)
	},
	\mkern25mu
	u \in \mathcal{U}
	\label{eq:TotalUtilityProblem_w_SINR}
	\end{align}
\end{subequations}
where $B_{u}\left(\mathcal{W} \right) = A_{u}\left({\bf 1}, \mathcal{W} \right)$.  
We use the Lagrangian for the equality constraints in~\eqref{eq:TotalUtilityProblem_w_SINR}
\vspace{-0.3em}
\begin{align}\label{eq:Lagrangian}
\mathcal{L}(\mathcal{W}, {\bm \gamma}, {\bm \nu})
&= \sum_{u \in \mathcal{U}} \delta_{u}
\log\left( 1 + 
\gamma_{u}
\right)
\nonumber \\
\noalign{\vskip-10pt}
&\quad
- \sum_{u \in \mathcal{U}} \nu_u \Bigg( \gamma_{u} - \frac{ 
	\sum_{r \in \mathcal{C}_u}
	\left|{\bf \hat{h}}_{ru}^H {\bf w}_{ru} \right|^2 }
{
	B_{u}\left( \mathcal{W} \right)} \Bigg)
\end{align}
When $\mathcal{W}$ is fixed, we evaluate the first optimality condition of the SINR auxiliary variable $\gamma_{u}$ by setting the derivative of \eqref{eq:Lagrangian} with respect to $\gamma_{u}$ to zero, which results in a value for $\nu_u$ that satisfies this optimality. Substituting $\nu_u$ back into~\eqref{eq:Lagrangian}:
\vspace{-0.3em}
\begin{align}\label{eq:objectiveFLag}
&
f_1(\mathcal{W}, {\bm \gamma})
=
\sum_{u \in \mathcal{U}} \delta_{u}
\left(
\log\left( 1 + 
\gamma_{u}
\right)
- \gamma_{u}
\right)
\nonumber \\
\noalign{\vskip-5pt}
&
+
\resizebox{0.41\textwidth}{!}
{$\displaystyle
	\sum_{u \in \mathcal{U}}
	\delta_{u}
	\Bigg( \frac{ \left(1 + \gamma_{u}\right) \sum_{r \in \mathcal{C}_u}
		\left|{\bf \hat{h}}_{ru}^H {\bf w}_{ru} \right|^2 }
	{ \displaystyle
		\sum_{r' \in \mathcal{B}} \sum_{u' \in \mathcal{E}_{r'}}  {\bf w}_{r'u'}^H  \left({\bf \hat{h}}_{r'u} {\bf \hat{h}}_{r'u}^H + {\bm \Theta_{r'u}} \right) {\bf w}_{r'u'}
		+ \sigma_z^2
	}
	\Bigg)
	$}
\end{align}
Setting the derivative of~\eqref{eq:objectiveFLag} to zero, we obtain the expected optimal formula for $\gamma_{u}$ in~\eqref{eq:TotalUtilityProblem_w_SINR}, which means they are equivalent.


Hence, our new reformulated problem can be written as
\vspace{-0.3em}
\begin{eqnarray}\label{eq:TotalProblem_lag_problem}
\stepcounter{probNum}
(\mathrm{P}\arabic{probNum})\quad
\max_{ \mathcal{W}, {\bm \gamma}}\quad & 
f_1( \mathcal{W}, {\bm \gamma})
& 
\\
&
\text{s.t.}\quad \eqref{eq:TotalUtilityProblem_w_indc1}~\mathrm{and}~\eqref{eq:TotalUtilityProblem_w_powerBudget} \nonumber
\end{eqnarray}
Note that we are not writing the dual problem here, but rather we are introducing SINR auxiliary variables ${\bm \gamma}$ that act as a proxy to account for the changes of the other variables.

\begin{prop}\label{prop:reordering_terms}
	Maximizing the second term in the objective function in~\eqref{eq:TotalProblem_lag_problem} is equivalent to maximizing the resulting $|\mathcal{C}_u|$ terms if we expand the numerator, where $|\mathcal{C}_u|$ is the size of the serving cluster for user $u$, i.e., the number of possible serving RRHs. If we decouple these terms and reorganize them with respect to each RRH $r$, we can restructure~\eqref{eq:objectiveFLag} as
	\begin{align}
	f_1(\mathcal{W}, {\bm \gamma})
	=
	\sum_{u \in \mathcal{U}} \delta_{u}
	\left(
	\log\left( 1 + 
	\gamma_{u}
	\right)
	- \gamma_{u}
	\right)
	+ \sum_{r \in \mathcal{B}}
	f_2(r; \mathcal{W}, {\bm \gamma}),
	\end{align}
	where for each RRH $r$ we have
	\begin{align}\label{eq:objectiveFLag_c_term}
	&f_2(r; \mathcal{W}, {\bm \gamma})
	=
	\nonumber \\
	&\ 
	\resizebox{0.42\textwidth}{!}
	{$\displaystyle
		\sum_{u \in \mathcal{E}_r}
		\delta_{u}
		\Bigg( \frac{ \left(1 + \gamma_{u}\right)
			\left|{\bf \hat{h}}_{ru}^H {\bf w}_{ru} \right|^2 }
		{ \displaystyle
			\sum_{r' \in \mathcal{B}} \sum_{u' \in \mathcal{E}_{r'}}  {\bf w}_{r'u'}^H  \left({\bf \hat{h}}_{r'u} {\bf \hat{h}}_{r'u}^H + {\bm \Theta_{r'u}} \right) {\bf w}_{r'u'}
			+ \sigma_z^2}
		\Bigg)
		$}
	\end{align}
	This restructuring follows from the fact that $\sum_{u \in \mathcal{U}} \left(\frac{ a_u \sum_{r \in \mathcal{C}_u} A_{ru}}{B_{u}} \right) = \sum_{r \in \mathcal{B}} \sum_{u \in \mathcal{E}_r} \left(\frac{a_u A_{ru}}{B_{u}} \right)$, where each term in the summation in~\eqref{eq:objectiveFLag_c_term} is the fraction of the useful signal received at user $u$ from RRH $r$ over the total signals received at this user (including the useful signals).
	
	Using fractional programming~\cite[Corollary~1]{FR8310563} over~\eqref{eq:objectiveFLag_c_term}, we can define the following function.
	\begin{align}\label{eq:objectiveFLag_c_term_linearized}
	&f_3(r; \mathcal{W}, {\bm \gamma}, {\bm \beta}_{r})
	=
	\sum_{u \in \mathcal{E}_r}
	\Bigg(
	2 \text{Re}\left\{
	\beta_{ru}^{*}
	\sqrt{\delta_{u}\left( 1 + \gamma_{u}\right)}
	{\bf w}_{ru}^H {\bf \hat{h}}_{ru}
	\right\}
	\nonumber \\
	&
	-
	\resizebox{0.46\textwidth}{!}
	{$\displaystyle
		\left.
		|\beta_{ru}|^2
		\left(
		\sum_{r' \in \mathcal{B}} \sum_{u' \in \mathcal{E}_{r'}}  {\bf w}_{r'u'}^H  \left({\bf \hat{h}}_{r'u} {\bf \hat{h}}_{r'u}^H + {\bm \Theta_{r'u}} \right) {\bf w}_{r'u'}
		+ \sigma_z^2
		\right)
		\right)
		$}
	\end{align}
	where vector ${\bm \beta}_{r} \in \mathbb{C}^{|\mathcal{E}_r|}$ is introduced as a new auxiliary variable, and $\text{Re}\{\cdot\}$ is the real part. 
	The function~\eqref{eq:objectiveFLag_c_term_linearized} is concave in ${\bm \beta}_{r}$. Also, it can be shown to be equivalent to~\eqref{eq:objectiveFLag_c_term} in the same way as was done with~\eqref{eq:objectiveFLag}, i.e., by setting the partial derivative with respect to $\beta_{ru}^{*}$ to zero, then substituting the value of $\beta_{ru}$ in~\eqref{eq:objectiveFLag_c_term_linearized} which yields~\eqref{eq:objectiveFLag_c_term}.
	%
	
	Then, our objective function in~\eqref{eq:TotalProblem_lag_problem} can be written as
	\begin{align}\label{eq:objectiveFLag_V2_withoutpower}
	f_4( \mathcal{W}, {\bm \gamma}, {\bm \beta})
	=
	\resizebox{0.35\textwidth}{!}
	{$\displaystyle
		\sum_{u \in \mathcal{U}} \delta_{u}
		\left(
		\log\left( 1 + 
		\gamma_{u}
		\right)
		- \gamma_{u}
		\right)
		+ 
		\sum_{r \in \mathcal{B}} f_3(r; \mathcal{W}, {\bm \gamma}, {\bm \beta}_{r})
		$}
	,
	\end{align}
	where ${\bm \beta} = \left[ \left({\bm \beta}_{1}\right)^T \dots \left({\bm \beta}_{|\mathcal{B}|}\right)^T \right]$ is the concatenation of the auxiliary variables ${\bm \beta}_{r} \in \mathbb{C}^{|\mathcal{E}_r|}$ introduced in \eqref{eq:objectiveFLag_c_term_linearized} for each RRH $r$.
	\qed
\end{prop}

\section{Resource Allocation}\label{section:ResourceAllocation}
\subsection{Optimal Expressions}
When the variables other than ${\bm \beta}_{r}$ are fixed, the optimal value of the auxiliary variable $\beta_{ru}$ can be obtained from its corresponding first-order optimality condition from~\eqref{eq:objectiveFLag_V2_withoutpower}~as
\begin{align}\label{eq:beta_opt}
\beta_{ru}
=
\frac{
	\sqrt{\delta_{u}\left( 1 + \gamma_{u}\right)}
	{\bf w}_{ru}^H  {\bf \hat{h}}_{ru}
}
{
	\sum_{r' \in \mathcal{B}} \sum_{u' \in \mathcal{E}_{r'}}  {\bf w}_{r'u'}^H  \left({\bf \hat{h}}_{r'u} {\bf \hat{h}}_{r'u}^H + {\bm \Theta_{r'u}} \right) {\bf w}_{r'u'}
	+ \sigma_z^2
}
\end{align}

Similarly for the beamformers ${\bf w}_{ru}$, we can write the Lagrangian formulation using the new objective function~\eqref{eq:objectiveFLag_V2_withoutpower} and the constraints in~\eqref{eq:TotalProblem_lag_problem}, then evaluating the corresponding first-order optimality condition to write ${\bf w}_{ru}$ as
\begin{align}\label{eq:beamformer_opt_2}
{\bf w}_{ru}
&=
\resizebox{0.43\textwidth}{!}
{$\displaystyle
	\sqrt{\delta_{u}\left( 1 + \gamma_{u}\right)}
	\beta_{ru}^{*}
	\bigg( \sum_{r' \in \mathcal{B}}\sum_{u' \in \mathcal{E}_{r'}}
	|\beta_{r'u'}|^2
	\left(
	{\bf \hat{h}}_{ru'} {\bf \hat{h}}_{ru'}^H
	+ {\bm \Theta_{ru'}}
	\right)
	$}
\nonumber \\
& \quad
+ \left( \mu_r + \lambda_r
\alpha_{ru}\right) {\bf I}_{M} \bigg)^{-1}
{\bf h}_{ru}
\end{align}
where the Lagrangian multipliers  $\lambda_r \geq 0$ and $\mu_r \geq 0$ correspond to the capacity \eqref{eq:TotalUtilityProblem_w_indc1} and power \eqref{eq:TotalUtilityProblem_w_powerBudget} constraints. Importantly, both these constraints relate to the power used at RRH $r$, i.e., both cannot be tight simultaneously. From complementary slackness, therefore, one of these Lagrange multipliers, both corresponding to RRH $r$, must be zero.

Unfortunately, we do not know a priori which constraint will remain tight. As we will see in our algorithm section, we propose a heuristic that, at each iteration of the algorithm, checks for whether the capacity constraint is satisfied (allowing $\lambda_r = 0$); if it is not satisfied, we update set $\lambda_r$ to a small value and update $\mu_r$ using a bisection search to meet the power constraint. Our results show that after a few iterations, $\lambda_r$ always converges to zero; we will comment on this in the results section. 

\subsection{Optimization Algorithm}
\begin{algorithm}
	\SetAlgoLined
	\SetInd{0.1em}{1em}
	\caption{User-scheduling and resource allocation}
	\label{algortihm:w_using_weights}
	Initialize $\mathcal{W}$ and weights $\alpha_{ru}$ for \emph{all} users.\label{step:L_1_norm_weights_init_beams}\\
	\While{ \textbf{NOT} converged}{\label{step:Algo_terminate_2}
		Update ${\bm \gamma}$ using~\eqref{eq:TotalUtilityProblem_w_SINR}.\label{step:using_weights_gamma_beta}\\
		Update ${\bm \beta}$ using~\eqref{eq:beta_opt}.\\
		%
		%
		%
		%
		Update $\mathcal{W}$ using~\eqref{eq:beamformer_opt_2}.\\
		Update $\{\mu_r, \lambda_r : r \in \mathcal{B} \}$ as described using complementary slackness.\\
		Update weights ${\bm \alpha}$ using~\eqref{eq:weightsUpdate}.\label{step:using_weights_alpha}\\
	}
\end{algorithm}	
We construct Algorithm~\ref{algortihm:w_using_weights} to allocate the resources for the users. The algorithm initializes some variables (Step~\ref{step:L_1_norm_weights_init_beams}) (e.g., conjugate beamforming to initialize ${\bf w}_{ru}$). Then, it updates the variables ${\bm \gamma}$, ${\bm \beta}$,  $\mathcal{W}$, and ${\bm \alpha}$ iteratively one at a time until convergence.

The complexity of updating ${\bm \gamma}$, ${\bm \beta}$, and ${\bm \alpha}$ is $\mathcal{O}\left(|\mathcal{U}|\right)$, $\mathcal{O}\left(|\mathcal{U}| C_\textrm{avg}\right)$, and $\mathcal{O}\left(|\mathcal{U}| C_\textrm{avg}\right)$, respectively, where $C_\textrm{avg}$ is the average cluster size per user, and it is affected by both the density of the active users and the large scale fading threshold $\rho$. The complexity of the beamforming using a weighted MMSE~\cite{WMMSE5756489} is $\mathcal{O}\left(|\mathcal{U}_s|^2 M^2 + |\mathcal{U}_s| M^3 \right)$, where $\mathcal{U}_s$ is the set of scheduled users. Hence, leading to a total algorithm complexity of at most $\mathcal{O}\left( M^3 |\mathcal{B}|^2 + M^4 |\mathcal{B}| + |\mathcal{U}|C_\textrm{avg} \right)$, where the number of the scheduled users is at most $|\mathcal{U}_s| \le M |\mathcal{B}|$.

\section{Numerical Results and Analysis}\label{section:results}
To eliminate network borders, we consider a wrap-around structure consisting of $Q = 7$ hexagonal \emph{virtual} cells\footnote{We create these virtual cells to allow for wrap-around; the cells have no physical meaning.} each having an inner radius $500~{\rm m}$ and containing $N$ RRHs that are uniformly distributed in each \emph{virtual} cell. Users are randomly distributed with a density $\lambda_\text{users}$ and a circular exclusion region of $20~{\rm m}$ around each RRH. We average our results using Monte Carlo simulations over both network realizations and time slots (TSs), and we include the effect of the users fairness by simulating $100$~TSs and averaging the results over the last allocated $50$~TSs, representing steady state performance\footnote{We emphasize that plotting the results from allocating a single TS would definitely give much higher performance, because the users with the best channel's conditions would be served, i.e., fairness is equal for all the users. Nonetheless, we are interested in studying the effect of the scheme on the long-term.}. 

We use the COST231 Walfish-Ikegami~\cite{Walfisch14401} to model the path loss at $1800$ MHz, resulting in $L_{|\mathrm{dB}}(d_{ru}) = - 112.4271 - 38\log_{10}\left(d_{ru}\right)$ where $d_{ru}$ is in $\mathrm{km}$. In Table~\ref{table:sim_parameters}, we summarize the parameters used.
\begin{table}[t]
	\centering
	\caption{Simulation parameters.}
	\begin{tabular}{|p{0.3\linewidth}|p{0.22\linewidth}|p{0.32\linewidth}|}
		\hline
		\hline
		& \multicolumn{1}{l}{ \textit{\textbf{Parameter}}} & \multicolumn{1}{|l|}{\textit{\textbf{Value}}}\\
		\hline
		Cell config. & $Q$, $N$, $M$, $\lambda_\text{users}$ & $7$, $10$, $8$, $200$ users/$\text{km}^2$\\
		\hline
		Power, Imperfect CSI & $p$, $\tau_d$, $(\tau_p)$, $p_u$  & $30~{\rm dBm}$,  $200$, $(16,\ 32,\ 64)$, $20~{\rm dBm}$\\
		\hline
		Noise spectral density, Noise figure & $S_z$, $F_z$, Bandwidth & $-174 {\rm dBm/Hz}$, $8~{\rm dBm}$, $180~{\rm KHz}$\\
		\hline
		Others & 
		$\sigma_\text{shadowing}$, $\rho$, $\eta$, $\epsilon$ & 
		$4~{\rm dB}$, $L(0.4)$, $0.2$, $\frac{0.9p}{M}$\\
		\hline
		\hline
	\end{tabular}
	\label{table:sim_parameters}
	\vspace{-1em}   
\end{table}

The proportional fairness weight, $\delta_{u}$, for user $u$ is the inverse of the achieved long-term average rate over an exponentially decaying window; in time slot $t$ we set $\delta_u$ as~\cite{yu2011adaptive}
\vspace{-0.3em}
\begin{align}
\delta_{u}^{(t)} = \frac{1}{\bar{R}_u^{(t)}}, \label{eq:UserWeight}
\end{align}
where $\delta_{u}^{(t)}$ is the value of $\delta_{u}$ at time slot $t$, and $\bar{R}_u^{(t)}$ is the user exponentially weighted rate averaged over previous time slots, and it is updated as $\bar{R}_u^{(t)} = \eta R_u^{(t)} + (1 - \eta) \bar{R}_u^{(t-1)}$ with a forgetting factor $\ 0 \le \eta \le 1$, where $R_u^{(t)}$ is the rate achieved by user $u$ at time~$t$, and it can be defined as~\cite{NoncoherentCRAN8482453}
\begin{align}\label{eq:ActualRate}
R_u^{(t)} =
\resizebox{0.41\textwidth}{!}
{$\displaystyle
\frac{\left(\tau_d - \tau_p\right)} {\tau_d}
\log\left(1 + \frac{ 
	\sum_{r \in \mathcal{C}_u} {s}_{ru}
	\left| {\bf h}_{ru}^H {\bf w}_{ru} \right|^2 }
{\displaystyle
	\sum_{r' \in \mathcal{B}} \sum_{u' \in \mathcal{E}_{r'},u' \ne u} {s}_{r'u'} \left| {\bf h}_{r'u}^H {\bf w}_{r'u'} \right|^2
	+ \sigma_z^2
} \right)
$}
\end{align}

\begin{figure}[t]
	\centering
	\begin{subfigure}{.485\columnwidth}
		\centering
		\includegraphics[width=1\textwidth]{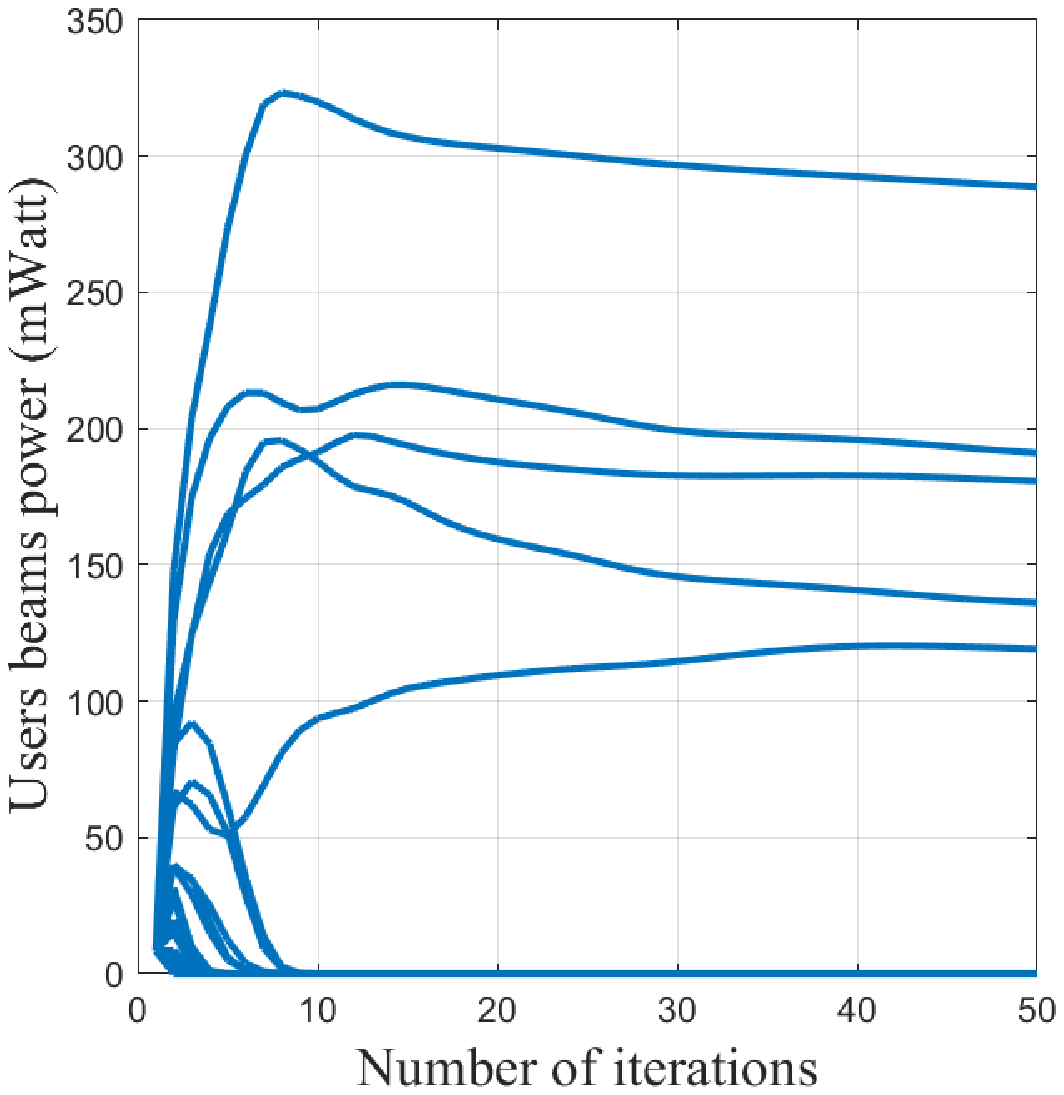}
		\caption{Allocated power for the users' beamformers on a typical RRH.}
		\label{fig:avgSpecEff1}
	\end{subfigure}%
	$\ $
	\begin{subfigure}{.485\columnwidth}
		\centering
		\includegraphics[width=1\textwidth]{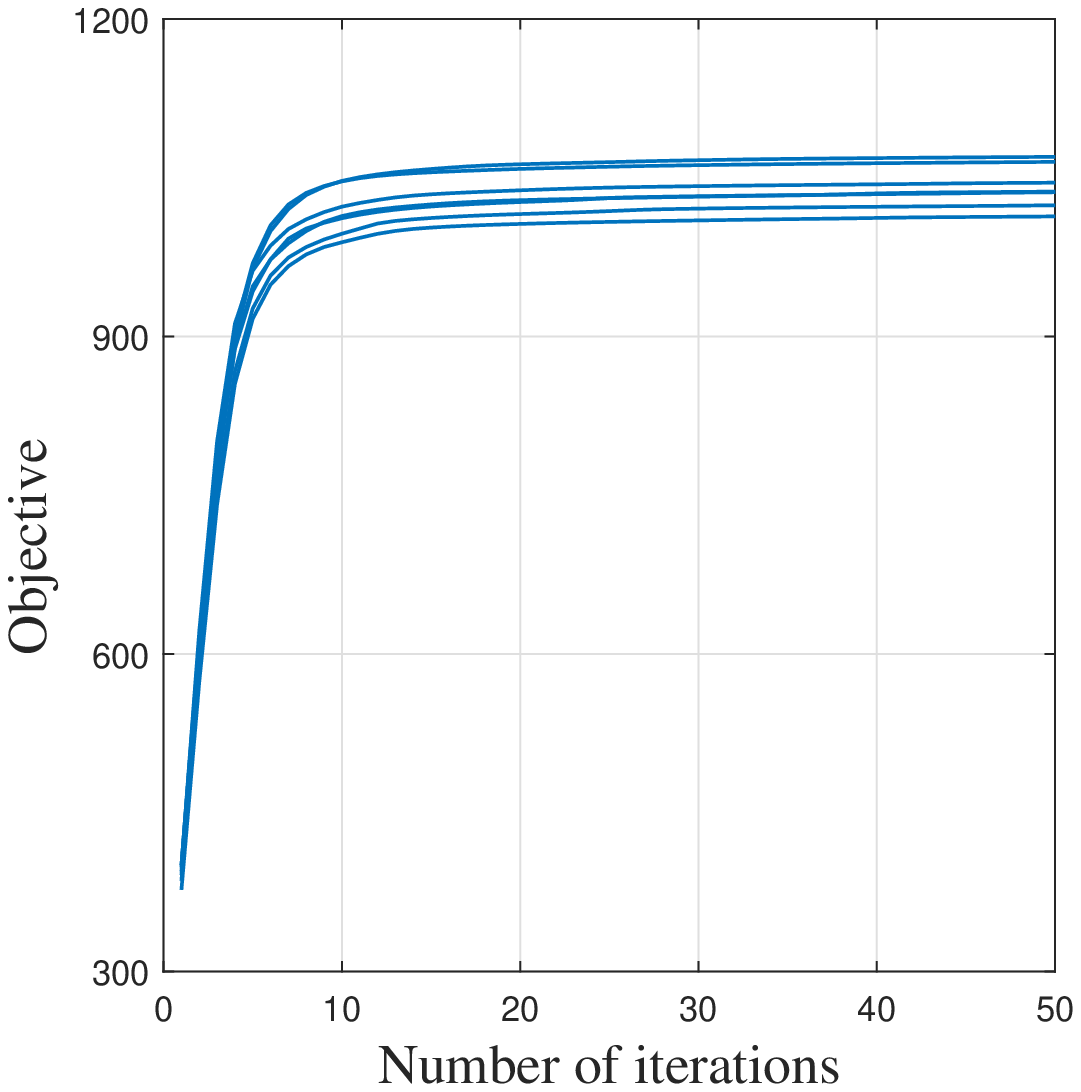}
		\caption{Convergence plot for many channel realizations.}
		\label{fig:convergencePlot_conf}
	\end{subfigure}%
	\caption{Evolution of the algorithm.}
	\label{fig:AlgoEvol}
\end{figure}

In Fig.~\ref{fig:AlgoEvol}(\subref{fig:avgSpecEff1}), we plot the evolution of the allocated power of the beamformer's weights for a typical RRH in a typical network as a function of the algorithm iterations. It is clear that after a few iterations, the power constraint~\eqref{eq:TotalUtilityProblem_w_powerBudget} is tighter than the capacity constraint~\eqref{eq:TotalUtilityProblem_w_indc1} which becomes deactivated, as previously discussed. In Fig.~\ref{fig:AlgoEvol}(\subref{fig:convergencePlot_conf}), we illustrate the convergence of the algorithm for several channel realizations.

\begin{figure}[H]
	\begin{subfigure}{.5\columnwidth}
		\centering
		\includegraphics[width=1\textwidth]{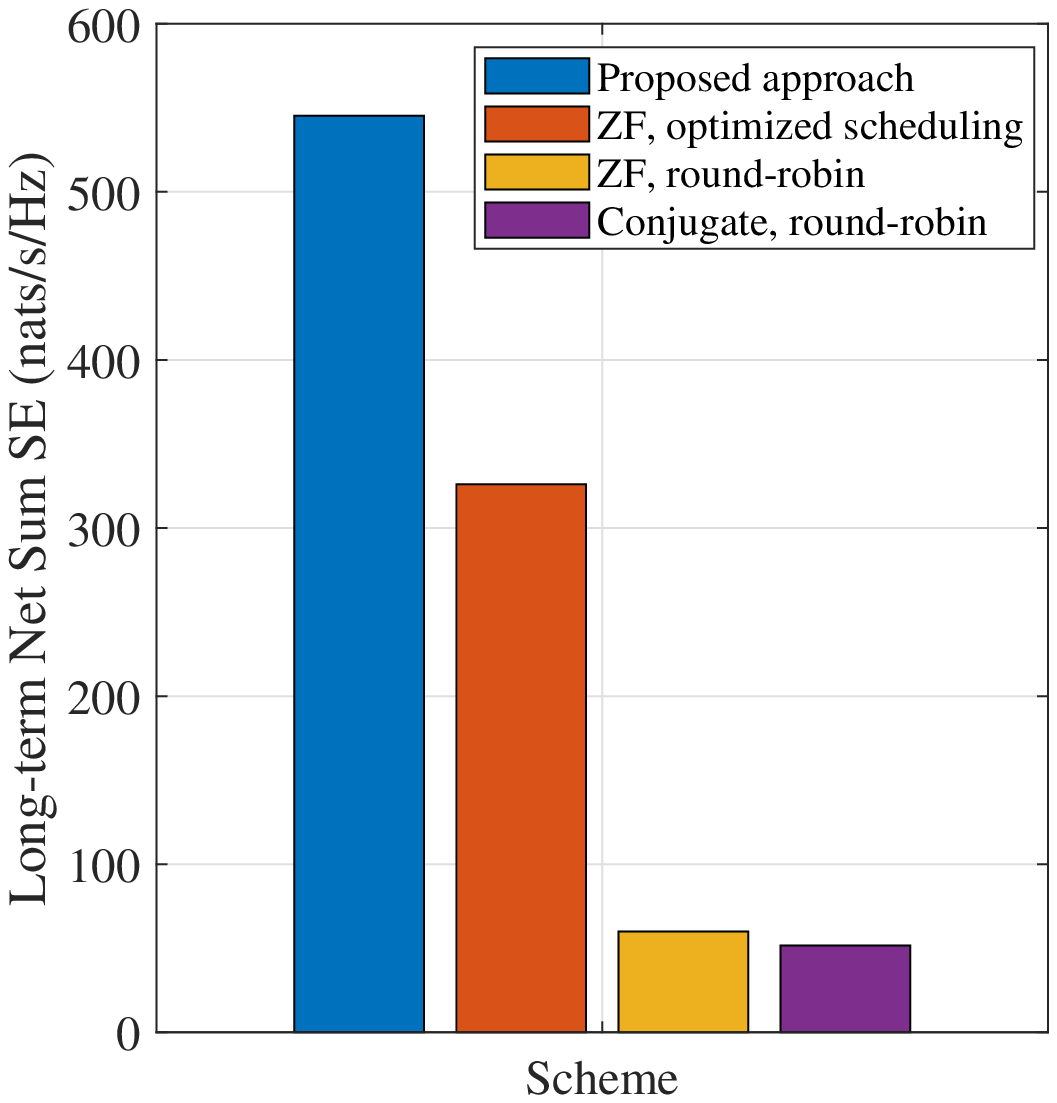}
		\caption{Network sum SE.}
		\label{sfig:NetSpectralEfficiency}
	\end{subfigure}%
	\begin{subfigure}{.49\columnwidth}
		\centering
		\includegraphics[width=1\textwidth]{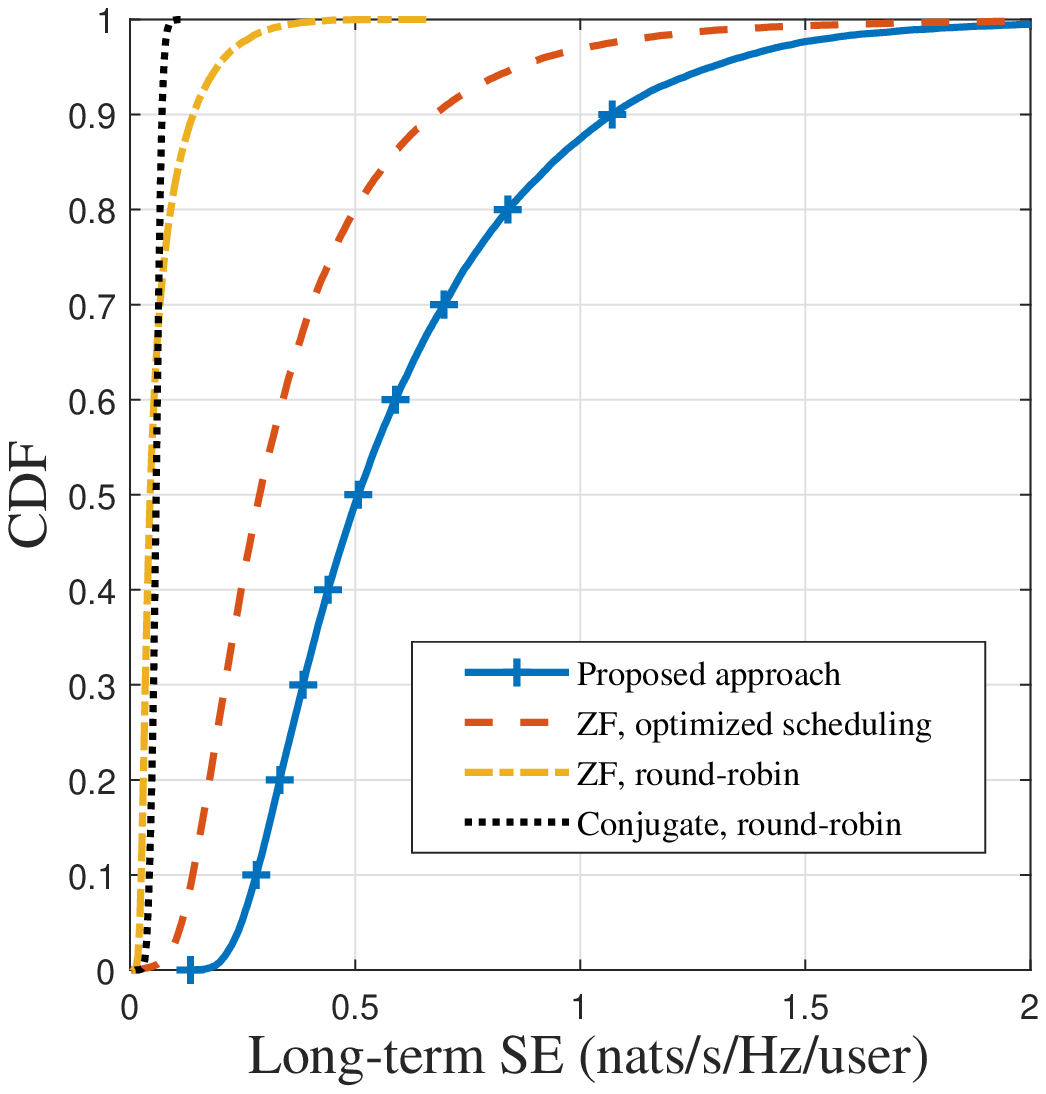}
		%
		%
		\caption{CDF of SE per user.}
		\label{sfig:CDFspectralEfficiency}
	\end{subfigure}%
	\caption{Long-term results, $N=10$.}
	\label{fig:LongTermIdeal}
\end{figure}

In Fig.~\ref{fig:LongTermIdeal}, we plot the long-term performance results under ideal CSI, i.e., the channels are known and there is no pilot training overhead. Fig.~\ref{fig:LongTermIdeal}(\subref{sfig:NetSpectralEfficiency}) shows the long-term network sum of spectral efficiency (SE) as a function of the algorithm iterations. The evolution of the curve shows that the algorithm converges smoothly with a non-decreasing fashion. Also, the results show a huge performance gain from using our approach compared to the ZF and the conjugate beamforming schemes with a round-robin scheduling. Compared to these schemes respectively, we obtain about a \myResultOne-fold and \myResultTwo-fold improvement. Additionally, we plot the resulting network sum SE when using the ZF beamforming scheme with the optimized user-scheduling obtained from our proposed approach. The results still show a \myResultThree-fold improvement. Clearly, this gap is due to the fact that the ZF beamformers are constructed at each RRH using only the channels of the served users, and each user is being allocated equal power irrespective of the channel conditions. Moreover, to quantify the effect of optimized user-scheduling, we compare the round-robin scheduling with that of the optimized one for the ZF beamforming. The result highlights the importance of optimized scheduling, where a \myResultFour-fold improvement is~achieved.

\vspace{-0.5em}

In Fig.~\ref{fig:LongTermIdeal}(\subref{sfig:CDFspectralEfficiency}), we plot the cumulative density function (CDF) of the long-term SE of the users, where an \myResultSELow\ to \myResultSEHigh-fold gain is observed in the median long-term user spectral efficiency for our approach compared to round-robin scheduling. We emphasize that this plot presents the long-term average rate that accounts for the user scheduling. Users may not be scheduled in every time slot; this is determined by their channels and their weights as defined in~\eqref{eq:UserWeight}. The gains for 
the 10$^{th}$-percentile rate, is clear. The proposed approach results in about $7$-fold and $2$-fold improvement in the cell-edge long-term rate compared to round-robin scheduling and ZF beamforming with optimized scheduling respectively.

To quantify the performance of imperfect CSI, we compare the following cases:
\begin{itemize}
	\item $\mathrm{PI}$: Our \emph{proposed} approach using \emph{ideal} channels, where no channel estimation phase is accounted for.
	\item $\mathrm{PEAR}$: Our \emph{proposed} approach when the algorithm is using the \emph{estimated} channel and using \emph{robust} beamforming, i.e., accounting for the estimation error. However, when plotting the results, we plot the \emph{actual} network performance, i.e., using~\eqref{eq:ActualRate}.
\end{itemize}

Since we use user-centric clustering, we define an \emph{area-based} pilot-reuse factor (not cell-based) as $\xi_p \triangleq \tau_p / \lambda_\text{users}$. For example, $\xi_p = 0.25$, means that on-average one-quarter of the users found in an area of $1\times 1\ {\rm km}^2$ are using orthogonal pilots. Under the user density specified in Table~\ref{table:sim_parameters}, the pilot sequence lengths $\tau_p = 64,\ 32,\ 16$ produce on-average $\xi_p = 0.32,\ 0.16,\ 0.08$ respectively.

\begin{figure}[H]
	\centering
	\includegraphics[width=0.75\columnwidth]{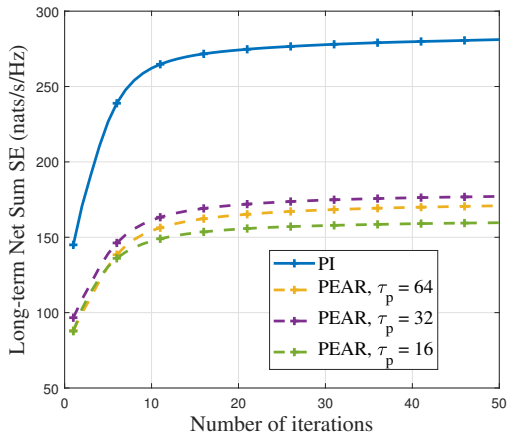}
	\caption{Evolution of the Long-term sum of SE, $N = 5$.}
	\label{fig:NetSpectralEfficiency_imperfect_robust}
\end{figure}

In Fig.~\ref{fig:NetSpectralEfficiency_imperfect_robust}, we plot the long-term network sum SE of the different studied cases using, but this time using $N = 5$ RRHs per virtual cell. The results show a drop of performance for $\mathrm{PEAR}$ by \myResultReuseFactorPerfRobust, \myResultReuseFactorPerfRobustTwo, and \myResultReuseFactorPerfRobustThree\ percent compared to the ideal channel case ($\mathrm{PI}$). If we are to quantify the performance drop due to only imperfect CSI, we obtain \myResultReuseFactorPerfRobustNF, \myResultReuseFactorPerfRobustNFTwo, and \myResultReuseFactorPerfRobustNFThree\ percent drop in the performance compared to the ideal case. From the results, using $\tau_p = 32$ provides the highest sum SE, i.e., it is a good compromise between the pilot contamination and the pilot-training overhead.

\vspace{-1em}
\section{Conclusion}\label{section:conclusion}
This paper optimized user-scheduling and resource allocation in a distributed cell-free MIMO system under the user-centric clustering scheme and the non-coherent transmission mode using a weighted sum rate problem formulation. We used tools from block coordinate descent, fractional programming, and compressive sensing to provide closed-form expressions for the optimized variables, while keeping the other variables fixed. This allowed us to construct an iterative optimization algorithm that converges smoothly in non-decreasing fashion. Our key contribution is optimized user-scheduling, which is neglected in most of the literature. The  numerical results show that our optimized resource allocation boosts network performance, both in terms of sum-rate and long-term proportional fair rates, compared to conventional round-robin schemes, where an \myResultSELow\ to \myResultSEHigh-fold gain in the long-term user spectral efficiency is observed.
\section*{Acknowledgment}
This work was supported in part by Ericsson Canada and in part by the Natural Sciences and Engineering Research Council (NSERC) of Canada.

\bibliography{RA_UserCentric_References}

\begin{thebibliography}{10}

\bibitem{PDPUsercentricVsDisjoint8969384}
H.~A. {Ammar} and R.~{Adve}, ``Power delay profile in coordinated distributed
  networks: User-centric v/s disjoint clustering,'' in {\em 2019 IEEE Global
  Conf. on Signal and Inf. Processing}, pp.~1--5, Nov 2019.

\bibitem{cellFreeVersusSmallCells7827017}
H.~Q. {Ngo}, A.~{Ashikhmin}, H.~{Yang}, E.~G. {Larsson}, and T.~L. {Marzetta},
  ``Cell-free massive {MIMO} versus small cells,'' {\em IEEE Transactions on
  Wireless Communications}, vol.~16, no.~3, pp.~1834--1850, 2017.

\bibitem{cellFreeUserCentricPower8901451}
S.~{Buzzi}, C.~{D’Andrea}, A.~{Zappone}, and C.~{D’Elia}, ``User-centric
  {5G} cellular networks: Resource allocation and comparison with the cell-free
  massive {MIMO} approach,'' {\em IEEE Transactions on Wireless
  Communications}, vol.~19, pp.~1250--1264, Feb 2020.

\bibitem{PrecodingDistrib2020Atzeni}
I.~{Atzeni}, B.~{Gouda}, and A.~{Tölli}, ``Distributed precoding design via
  over-the-air signaling for cell-free massive {MIMO},'' {\em IEEE Transactions
  on Wireless Communications}, pp.~1--1, 2020.

\bibitem{6920005}
B.~{Dai} and W.~{Yu}, ``Sparse beamforming and user-centric clustering for
  downlink cloud radio access network,'' {\em IEEE Access}, vol.~2,
  pp.~1326--1339, 2014.

\bibitem{maxMinRate8756286}
M.~{Bashar}, K.~{Cumanan}, A.~G. {Burr}, H.~Q. {Ngo}, M.~{Debbah}, and
  P.~{Xiao}, ``Max–min rate of cell-free massive {MIMO} uplink with optimal
  uniform quantization,'' {\em IEEE Transactions on Communications}, vol.~67,
  no.~10, pp.~6796--6815, 2019.

\bibitem{powerControlCellFree7917284}
E.~{Nayebi}, A.~{Ashikhmin}, T.~L. {Marzetta}, H.~{Yang}, and B.~D. {Rao},
  ``Precoding and power optimization in cell-free massive {MIMO} systems,''
  {\em IEEE Trans. on Wireless Comm.}, vol.~16, pp.~4445--4459, July 2017.

\bibitem{9107496}
Q.~D. {Vu}, L.~N. {Tran}, and M.~{Juntti}, ``Noncoherent joint transmission
  beamforming for dense small cell networks: Global optimality, efficient
  solution and distributed implementation,'' {\em IEEE Transactions on Wireless
  Communications}, vol.~19, no.~9, pp.~5891--5907, 2020.

\bibitem{ResourceAllo6175089}
D.~W.~K. {Ng}, E.~S. {Lo}, and R.~{Schober}, ``Dynamic resource allocation in
  {MIMO-OFDMA} systems with full-duplex and hybrid relaying,'' {\em IEEE
  Transactions on Communications}, vol.~60, no.~5, pp.~1291--1304, 2012.

\bibitem{ResourceAllo1658226}
W.~{Yu} and R.~{Lui}, ``Dual methods for nonconvex spectrum optimization of
  multicarrier systems,'' {\em IEEE Transactions on Communications}, vol.~54,
  no.~7, pp.~1310--1322, 2006.

\bibitem{FR8310563}
K.~{Shen} and W.~{Yu}, ``Fractional programming for communication
  systems—part {II}: Uplink scheduling via matching,'' {\em IEEE Transactions
  on Signal Processing}, vol.~66, pp.~2631--2644, May 2018.

\bibitem{Ahmad9084256}
A.~A. {Khan}, R.~S. {Adve}, and W.~{Yu}, ``Optimizing downlink resource
  allocation in multiuser {MIMO} networks via fractional programming and the
  hungarian algorithm,'' {\em IEEE Transactions on Wireless Communications},
  vol.~19, no.~8, pp.~5162--5175, 2020.

\bibitem{NoncoherentCRAN8482453}
C.~{Pan}, H.~{Ren}, M.~{Elkashlan}, A.~{Nallanathan}, and L.~{Hanzo}, ``The
  non-coherent ultra-dense {C-RAN} is capable of outperforming its coherent
  counterpart at a limited fronthaul capacity,'' {\em IEEE Journal on Selected
  Areas in Communications}, vol.~36, no.~11, pp.~2549--2560, 2018.

\bibitem{frenger2019antenna}
P.~Frenger, J.~Hederen, M.~Hessler, and G.~Interdonato, ``Antenna arrangement
  for distributed massive {MIMO},'' Nov.~28 2019.
\newblock US Patent App. 16/435,054.

\bibitem{kay1993fundamentals}
S.~M. Kay, {\em Fundamentals of Statistical Signal Processing, vol.~1:
  Estimation Theory}.
\newblock Prentice Hall PTR, 1993.

\bibitem{karypis2000comparison}
M.~G. Karypis, V.~Kumar, and M.~Steinbach, ``A comparison of document
  clustering techniques,'' in {\em TextMining Workshop at KDD2000 (May 2000)},
  2000.

\bibitem{candes2008enhancing}
E.~J. {Candes}, M.~B. {Wakin}, and S.~P. {Boyd}, ``Enhancing sparsity by
  reweighted $\ell_1$ minimization,'' {\em Journal of Fourier analysis and
  applications}, vol.~14, no.~5-6, pp.~877--905, 2008.

\bibitem{WMMSE5756489}
Q.~{Shi}, M.~{Razaviyayn}, Z.~{Luo}, and C.~{He}, ``An iteratively weighted
  {MMSE} approach to distributed sum-utility maximization for a {MIMO}
  interfering broadcast channel,'' {\em IEEE Transactions on Signal
  Processing}, vol.~59, pp.~4331--4340, Sep. 2011.

\bibitem{Walfisch14401}
J.~{Walfisch} and H.~L. {Bertoni}, ``A theoretical model of {UHF} propagation
  in urban environments,'' {\em IEEE Transactions on Antennas and Propagation},
  vol.~36, pp.~1788--1796, Dec 1988.

\bibitem{yu2011adaptive}
W.~Yu, T.~Kwon, and C.~Shin, {\em Adaptive resource allocation in cooperative
  cellular networks}, ch.~9, pp.~233--256.
\newblock Cambridge Univ. P., 2011.

\end{thebibliography}
\bibliographystyle{ieeetr}

\end{document}